\documentclass{PoS}

\PoS{PoS(HEP2005)093}

\def\Dzkpi{D^0\to K^-\pi^+}
\def\Dzkpipiz{D^0\to K^-\pi^+\pi^0}
\def\Dzkpipipi{D^0\to K^-\pi^+\pi^+\pi^-}
\def\Dpkpipi{D^+\to K^-\pi^+\pi^+}
\def\Dpkpipipiz{D^+\to K^-\pi^+\pi^+\pi^0}
\def\Dpkspi{D^+\to K^0_S\pi^+}
\def\Dpkspipiz{D^+\to K^0_S\pi^+\pi^0}
\def\Dpkspipipi{D^+\to K^0_S\pi^+\pi^+\pi^-}
\def\Dpkkpi{D^+\to K^+K^-\pi^+}

\def\Dzbarkpi{\bar D^0\to K^+\pi^-}
\def\Dzbarkpipiz{\bar D^0\to K^+\pi^-\pi^0}
\def\Dzbarkpipipi{\bar D^0\to K^+\pi^-\pi^-\pi^+}
\def\Dmkpipi{D^-\to K^+\pi^-\pi^-}
\def\Dmkpipipiz{D^-\to K^+\pi^-\pi^-\pi^0}
\def\Dmkspi{D^-\to K^0_S\pi^-}
\def\Dmkspipiz{D^-\to K^0_S\pi^-\pi^0}
\def\Dmkspipipi{D^-\to K^0_S\pi^-\pi^-\pi^+}

\def\NDDbar{N_{D\bar D}}
\def\NDzDzbar{N_{D^0\bar D^0}}
\def\NDpDm{N_{D^+D^-}}

\title{Hadronic Branching Fractions \& $D\overline{D}$ Cross-sections at 
       $\psi$(3770) from CLEO-c}

\ShortTitle{Hadronic branching fractions \& DDbar xsect at CLEO}

\author{\speaker{Yongsheng Gao (for CLEO Collaboration)}\\

        Southern Methodist University, Dallas, Texas 75275-0175, USA\\

        E-mail: \email{gao@mail.physics.smu.edu}}

\abstract{
We present some recent results in hadronic decays
and cross-section measurements at $\psi$(3770) from CLEO Collaboration. 
They include measurement of absolute hadronic branching fractions of $D$
         mesons and $e^+e^-\to D\bar D$ cross sections,
inclusive production of $\eta$, $\eta^{,}$ and $\phi$ in $D$ decays,
branching fractions of $D^{+} \to K^{0}_{S,L} \pi^{+}$ and $\eta \pi^{+}$,
$\psi$(3770) non-$D\bar D$ decays, and
timelike electromagnetic form factors of pion, kaon, and proton.
These results are based on  55.8 pb$^{-1}$ and  281 pb$^{-1}$ at
$\psi$(3770), and other data samples collected by the CLEO-c detector at 
the Cornell Electron Storage Ring (CESR). 
}

\FullConference{International Europhysics Conference on High Energy Physics\\

                 July 21st - 27th 2005\\

                 Lisboa, Portugal}

\begin{document}

\section{Measurement of Absolute Hadronic Branching Fractions of $D$
         Mesons and $e^+e^-\to D\bar D$ Cross Sections at $E_{\rm cm}=3773$ MeV}

Absolute measurements of hadronic charm meson branching
fractions play a central role in the study of the weak
interaction because they serve to normalize many $D$ and $B$
meson branching fractions, from which elements of the
Cabibbo-Kobayashi-Maskawa (CKM) matrix~\cite{ckm} are determined.
For instance, the determination of the CKM matrix element $|V_{cb}|$ from the
$B\to D^*\ell\nu$ decay rate using full $D^*$ reconstruction requires 
knowledge of the
$D$ meson branching fractions~\cite{vcbreview,PDG}.
We present charge-averaged branching fraction measurements of
three $D^0$ and six $D^+$ decay modes~\cite{prl121801}.

The data sample consists of 55.8 ${\rm pb}^{-1}$ of integrated luminosity
collected by the CLEO-c detector on the $\psi(3770)$ resonance, at
a center-of-mass energy $E_{\rm cm}=3773$ MeV.
Reconstruction of one $D$ or
$\bar D$ meson (called single tag or ST) tags the event
as either $D^0\bar D^0$ or $D^+D^-$.  For a given decay mode $i$, we
measure independently the $D$ and $\bar D$ ST yields,
denoted by $N_i$ and $\bar N_i$. We determine the corresponding efficiencies,
denoted by $\epsilon_i$ and $\bar\epsilon_i$,
from Monte Carlo simulations.
Thus, $N_i=\epsilon_i{\cal B}_iN_{D\bar D}$ and
$\bar N_i=\bar\epsilon_i{\cal B}_i N_{D\bar D}$, 
where ${\cal B}_i$ is the branching fraction for
mode $i$,  assuming no $CP$ violation, and $N_{D\bar D}$ is the 
number of produced $D\bar D$ pairs.  Double tag (DT) events are the subset of ST
 events
where both the $D$ and $\bar D$ are reconstructed.  The DT
yield for $D$ mode $i$ and $\bar D$ mode $j$, denoted by
$N_{ij}$, is given by
$N_{ij} = \epsilon_{ij}{\cal B}_i{\cal B}_jN_{D\bar D}$,
where $\epsilon_{ij}$ is the DT efficiency.  As with ST yields,
the charge conjugate DT yields and efficiencies, $N_{ji}$ and
$\epsilon_{ji}$, are determined separately.  Charge conjugate
particles are implied, unless referring to ST and DT yields.

We extract branching fractions and $\NDDbar$ by combining
ST and DT yields with a least squares technique.
We fit $D^0$ and $D^+$ parameters simultaneously, including in the $\chi^2$
statistical and systematic uncertainties and their correlations for all
experimental inputs~\cite{brfit}.
We measure 9 ST and 45 DT yields in data and determine the
corresponding efficiencies from simulated events.

\begin{table}[tb]
\caption{Fitted branching fractions and $D\bar D$ pair yields, along with the
fractional FSR corrections.  Uncertainties are statistical and systematic,
respectively. The Particle Data Group~\protect{\cite{PDG}} lists the average 
branching fractions ${\cal B}(D^0\to K^-\pi^+)=(3.85\pm0.09)\%$ and
${\cal B}(D^+\to K^-\pi^+\pi^+)=(9.1\pm0.7)\%$.
}
\label{tab-dataResults}
\begin{center}
\begin{tabular}{lccc}
\hline\hline
Parameter & Fitted Value & ~$\Delta_{\rm FSR}$ \\
\hline
$\NDzDzbar$               & $(2.01\pm 0.04\pm 0.02)\times 10^5$ & ~$-0.2\%$ \\
${\cal B}(\Dzkpi)$        & $(3.91\pm 0.08\pm 0.09)\%$        & ~$-2.0\%$ \\
${\cal B}(\Dzkpipiz)$     & $(14.9\pm 0.3\pm 0.5)\%$           & ~$-0.8\%$ \\
${\cal B}(\Dzkpipipi)$    & $(8.3\pm 0.2\pm 0.3)\%$           & ~$-1.7\%$ \\
\hline
$\NDpDm$                  & $(1.56\pm 0.04\pm 0.01)\times 10^5$ & ~$-0.2\%$ \\
${\cal B}(\Dpkpipi)$      & $(9.5\pm 0.2\pm 0.3)\%$           & ~$-2.2\%$ \\
${\cal B}(\Dpkpipipiz)$   & $(6.0\pm 0.2\pm 0.2)\%$           & ~$-0.6\%$ \\
${\cal B}(\Dpkspi)$       & $(1.55\pm 0.05\pm 0.06)\%$        & ~$-1.8\%$ \\
${\cal B}(\Dpkspipiz)$    & $(7.2\pm 0.2\pm 0.4)\%$           & ~$-0.8\%$ \\
${\cal B}(\Dpkspipipi)$   & $(3.2\pm 0.1\pm 0.2)\%$           & ~$-1.4\%$ \\
${\cal B}(\Dpkkpi)$       & $(0.97\pm 0.04\pm 0.04)\%$        & ~$-0.9\%$ \\
\hline\hline
\end{tabular}
\end{center}
\end{table}

\begin{table}[tb]
\caption{Ratios of branching fractions to the reference
branching fractions ${\cal R}_0\equiv {\cal B}(\Dzkpi)$ and
${\cal R}_\pm\equiv {\cal B}(\Dpkpipi)$, along with the
fractional FSR corrections.
Uncertainties are statistical and systematic, respectively.}
\label{tab-dataResultsRatios}
\begin{center}
\begin{tabular}{lcc}
\hline\hline
Parameter & Fitted Value & ~$\Delta_{\rm FSR}$ \\
\hline
${\cal B}(\Dzkpipiz)/{\cal R}_0$ & $3.65\pm 0.05\pm 0.11$ & ~$+1.2\%$ \\
${\cal B}(\Dzkpipipi)/{\cal R}_0$ & $2.10\pm 0.03\pm 0.06$ & ~$+0.3\%$ \\
\hline
${\cal B}(\Dpkpipipiz)/{\cal R}_\pm$  & $0.613\pm 0.013\pm 0.019$ & ~$+1.7\%$\\
${\cal B}(\Dpkspi)/{\cal R}_\pm$ & $0.165\pm 0.004\pm 0.006$ & ~$+0.4\%$ \\
${\cal B}(\Dpkspipiz)/{\cal R}_\pm$ & $0.752\pm 0.016\pm 0.033$ & ~$+1.4\%$ \\
${\cal B}(\Dpkspipipi)/{\cal R}_\pm$ & $0.340\pm 0.009\pm 0.014$ & ~$+0.8\%$ \\
${\cal B}(\Dpkkpi)/{\cal R}_\pm$ & $0.101\pm 0.004\pm 0.002$ & ~$+1.3\%$ \\
\hline\hline
\end{tabular}
\end{center}
\end{table}

The results of the data fit are shown in Table~\ref{tab-dataResults}.
We also compute ratios of branching fractions to the reference
branching fractions, shown in Table~\ref{tab-dataResultsRatios}.  These
ratios have higher precision than the individual
branching fractions, and they also agree with the PDG averages.  
We obtain the $e^+e^-\to D\bar D$ cross sections by scaling
$\NDzDzbar$ and $\NDpDm$ by the luminosity, which we determine to be
${\cal L} = (55.8 \pm 0.6)$ ${\rm pb}^{-1}$.  Thus, at
$E_{\rm cm}=3773$ MeV, we find peak cross sections of
$\sigma( e^+e^-\to D^0\bar D^0 ) = (3.60\pm 0.07^{+0.07}_{-0.05}) \ {\rm nb}$,
$\sigma( e^+e^-\to D^+ D^- ) = (2.79\pm 0.07^{+0.10}_{-0.04}) \ {\rm nb}$,
$\sigma( e^+e^-\to D\bar D ) = (6.39\pm 0.10^{+0.17}_{-0.08}) \ {\rm nb}$, and
$\sigma( e^+e^-\to D^+ D^- ) / \sigma( e^+e^-\to D^0\bar D^0 ) = 0.776\pm 0.024^
{+0.014}_{-0.006}$,
where the uncertainties are statistical and systematic, respectively.

\section{Inclusive Production of $\eta$, $\eta^{,}$ and $\phi$ in $D$ Decays}

Using 281 ${\rm pb}^{-1}$ of full CLEO-c data at the $\psi(3770)$ resonance, 
we measure~\cite{CLEOCONF05-4} inclusive production of $\eta$, $\eta^{,}$ and 
$\phi$ in $D$ Decays.
The tag yields are shown in Table~\ref{tagyield-281}. The measured inclusive 
branching fractions of $\eta$, $\eta^{,}$ and $\phi$ in $D$ Decays, together
with comparison with PDG, are shown in Table~\ref{inclusive-result}. They
represent significant improvements comparing with current PDG measurements.

\begin{table}
\caption{Single tag data yields and efficiencies and their
         background from the 281 pb$^{-1}$ data sample.}
\label{tagyield-281}
\begin{center}
\begin{tabular}{lcc}
\hline\hline
$D$ Tag Mode           &  Yield           & Background \\ \hline
$\Dzbarkpi$            &  $ 49418\pm 246$  &      $630$ \\
$\Dzbarkpipiz$         &  $101960\pm 476$  &    $18307$ \\
$\Dzbarkpipipi$        &  $ 76178\pm 306$  &     $6421$ \\ \hline
$\Dmkpipi$             &  $ 77387\pm 281$  &     $1868$ \\
$\Dmkpipipiz$          &  $ 24850\pm 214$  &    $12825$ \\
$\Dmkspi$              &  $ 11162\pm 136$  &      $514$ \\
$\Dmkspipiz$           &  $ 20244\pm 170$  &      $170$ \\
$\Dmkspipipi$          &  $ 18176\pm 255$  &      $255$ \\
\hline\hline
\end{tabular}
\end{center}
\end{table}

\begin{table}
\caption{Results of inclusive production of $\eta$, $\eta^{,}$ and $\phi$ 
         in $D$ decays from 281 pb$^{-1}$ data sample.}
\label{inclusive-result}
\begin{center}
\begin{tabular}{lcccc}
\hline\hline
Mode         &  $D^{0}$ BR (\%)      & PDG (\%) 
             &  $D^{+}$ BR (\%)      & PDG (\%) \\ \hline
$\eta$X      &  $9.4\pm 0.4\pm 0.6$  & $<$13     
             &  $5.7\pm 0.5\pm 0.5$  & $<$13    \\
$\eta^{,}$X  &  $2.6\pm 0.2\pm 0.2$  &  $-$     
             &  $1.0\pm 0.2\pm 0.1$  &  $-$     \\
$\phi$X      &  $1.0\pm 0.1\pm 0.1$  & $1.7\pm 0.8$     
             &  $1.1\pm 0.1\pm 0.2$  & $<$1.8   \\
\hline\hline
\end{tabular}
\end{center}
\end{table}

\section{Measurements of $D^{+} \to K^{0}_{S,L} \pi^{+}$ and $\eta \pi^{+}$ Branching
         Fractions} 

Using the single charge $D$ tag from 281 pb$^{-1}$ of CLEO-c data at $\psi$(3770), 
we measure the branching fraction of $D^{+} \to K^{0}_{S,L} \pi^{+}$ using
a missing mass technique: 
${\cal B}(D^{+} \to K^{0}_{S,L} \pi^{+})$ =  $(3.06\pm 0.06\pm 0.16)$\%,
${\cal B}(D^{+} \to \eta \pi^{+})$        =  $(0.39\pm 0.03\pm 0.03)$\%.

\section{$\psi$(3770) non-$D\bar D$ Decays}

We also measure the branching fractions of $\psi$(3770) to non-$D\bar{D}$ final
states, including Vector-Pseudoscalar and multi-body final states. The details of
the event selections and results can be found in Ref.~\cite{CLEOCONF05-1,CLNS05-1921}.

\section{Precision Measurements of the Timelike Electromagnetic 
         Form Factors of Pion, Kaon, and Proton}

Using 20.7 pb$^{-1}$ of $e^+e^-$ annihilation data taken at $\sqrt{s}=3.671$ GeV
with the CLEO-c detector, we make precision measurements of the electromagnetic form 
factors of the charged pion, charged kaon, and proton for timelike 
momentum transfer of $|Q^2|=13.48$ GeV$^2$ by the reaction $e^+e^-\to h^+h^-$.  
The measurements~\cite{CLNS05-1936} are the first ever with identified pions 
and kaons of $|Q^2|>4$ GeV$^2$, with the results 
$F_\pi(13.48\;\mathrm{GeV}^2)=0.075\pm0.008(\mathrm{stat})\pm0.005(\mathrm{syst}
)$ 
and 
$F_K(13.48\;\mathrm{GeV}^2)=0.063\pm0.004(\mathrm{stat})\pm0.001(\mathrm{syst})$.  
The result for the proton, assuming $G^p_E=G^p_M$, is 
$G^p_M(13.48\;\mathrm{GeV}^2)=0.014\pm0.002(\mathrm{stat})\pm0.001(\mathrm{syst}
)$, 
which is in agreement with earlier results.

\end{document}